\documentclass[12pt,a4paper]{article}
\usepackage[utf8]{inputenc}
\usepackage{amsmath}
\usepackage{graphicx}
\usepackage{amssymb}
\usepackage{amsthm}

\usepackage[breaklinks=true,backref=page]{hyperref}

\usepackage[english]{babel}
\usepackage[square,numbers]{natbib}

\usepackage[nottoc]{tocbibind}

\usepackage{amsthm}
\theoremstyle{plain}
\newtheorem{thm}{Teorema}[section]

\newtheorem{prop}[thm]{Proposición}

\theoremstyle{definition}

\theoremstyle{remark}

\title{General model of sex distribution, mating probability and egg production for macroparasites with polygamous mating system}
\author{Gonzalo Maximiliano LOPEZ$^{1,3,4}$, Juan Pablo APARICIO$^{1,2}$\\
	\\
	{\small $^1$ Instituto de Investigaciones en Energ\'ia no Convencional (INENCO),} \\ {\small Consejo Nacional de Investigaciones Cient\'ificas y T\'ecnicas (CONICET),}\\
	{\small Universidad Nacional de Salta, Av. Bolivia 5150, 4400 Salta, Argentina.}\\
	$^2${\small Simon A. Levin Mathematical, Computational and Modeling Sciences Center,} \\ {\small Arizona State University, PO Box 871904 Tempe, AZ 85287-1904, USA}\\
	{\small $^3$ Departamento de Matem\'atica,}\\{\small Universidad Nacional de Salta, Av. Bolivia 5150, 4400 Salta, Argentina.}\\
	{\small $^4$ Corresponding author: gonzalo.maximiliano.lopez@gmail.com}}
	\date{}

\begin{document}
\maketitle
\tableofcontents

\begin{abstract}
%
%
	The reproductive habits of helminths are important for the study of the dynamics of their transmission.
	For populations of parasites distributed by Poisson or negative binomial models, these habits have already been studied. 
	However, there are other statistical models that describe these populations, such as zero-inflated models, but where reproductive characteristics were not analyzed. Using an arbitrary model for the parasite population, we model the distribution of females and males per host, and from these we model the different reproductive variables such as the mean number of fertile females, the mean egg production, the mating probability, the mean fertilized egg production.
	We show that these variables change due to the effects of a negative density-dependence fecundity, a characteristic of helminth parasites. We present the results obtained for some particular models.
	
\end{abstract}
	\section{Introduction}
	The most important factors in understanding the transmission dynamics of soil-transmitted helminths are reproductive behaviors.
	
	Most helminths that infect human are dioecious (separate sexes) and many are assumed to be polygamous (the presence of at least one male can guarantee the fertility of all females present), but quantitative data are not available\cite{anderson1992infectious}.
	
	The production of offspring of these parasites is, in general, a function of their population size, the proportion of females, and their reproductive behavior.
	Developing mathematical models that allow understanding the distribution by sex (female and male) and the reproductive behavior of these parasites is important.
	
	
	In a population where the distribution of parasites per host is described by a Poisson or negative binomial statistical model, the distribution by sex was studied for the case of a sex ratio 1:1 in \cite{may1977togetherness} and for a variable sex ratio in \cite{may1993biased}.
	Also a dynamic model for the number of fertilized females is presented in \cite{leyton1968stochastic}.
	
	
	
	In this work we present a generalization of what was developed by previous mentioned works.
	To model the distribution by sex, we will assume an arbitrary model for the distribution of parasites per host and variable sex ratios. We also consider that the distributions by sex can be constituted jointly or independently.

	
	We then calculated different reproductive variables such as mean number of fertile females, mean egg production, mating probability, and mean fertile egg production.
	Finally we show that these variables change with the density of parasites per host.
	

%
		
	\section{Distribution of parasites by sex}\label{sec:distsexo}
	For each individual parasite burden, the fraction of all females and males parasites are represented by $\alpha$ and $\beta$, respectively, where $\alpha+\beta=1$.
	Then the ratio of males to females is given by $\beta / \alpha : 1$. Also if $m$ is the mean of the distribution of parasites, the mean number of parasites
	females and males are given by $\alpha m$ and $\beta m $ respectively.

	Let $W$ be a random variable, the number of parasites per host and denoted by $F$ the number of female parasites per host.
	We propose that the distribution of females parasites per host is modeled by a stopped sums distribution (\cite{johnson2005univariate}) and its probability generating function (pgf) is the function $G_W \circ G_B$, where $G_B$  is the pgf of the Bernoulli distribution ($G_B(s)=\beta + \alpha s$)\cite{johnson2005univariate}. 
	Therefore the variable $F$ is given by $F=\sum_{i=1} ^W Y_i$ where $Y_i\sim \mathrm{Ber}(\alpha)$, and its pgf is

	\begin{equation}\label{genf}
	\begin{split}
	G_F(s)=&G_W(\beta + \alpha s)\\
	=&\sum_{w\geq 0}\sum_{j=0}^{w} \mathrm{Pr}(W=w)\binom{w}{j}\alpha^j\beta^{w-j}s^j
	\end{split}
	\end{equation}
	The first moments of $F$ are
	\begin{equation}
	\mu_F=\alpha \mu_W \qquad \sigma_F^2=\alpha^2\sigma_W^2+ \alpha\beta\mu_W
	\end{equation}
	The coefficient of dispersion, or variance-to-mean ratio 
	$D=\frac{\sigma_F^2}{\mu_F}$, is given by \[D=\alpha\frac{\sigma_W^2}{\mu_W}+\beta\]
	where $\frac{\sigma_W^2}{\mu_W}$ 
	is variance-to-mean ratio of $W$.
	Therefore, if $W$ has overdispersion, so will $F$. 
	
	
	Similarly, if $M$ is the number of male parasites, this variable is given by $M = W - F$ where its mean is $\mu_M=\beta\mu_W$. By the definition of $F$ and $M$ these are dependent variables.
	

	
	
	\section{Mating probability}\label{sec:probapareamiento}
	\subsection{Mean number of fertilized female parasites}

	The parasites treated in this work present a polygamous mating system, so the presence of at least one male parasite in the host ensures the fertility of all females.
	Therefore, from the distribution of parasites by sex presented in \eqref{genf}, the mean number of fertilized female parasites per host is given by
	\begin{equation}\label{eqhembrasfecun}
	\begin{split}
	\sum_{n\geq 1}\sum_{j=0}^{n-1}j p_n\binom{n}{j}\alpha^j\beta^{n-j}
	&=\alpha  m - \alpha G'(\alpha)
	\end{split}
	\end{equation}
	where the term $\sum_{j=0}^{n-1}j p_n\binom{n}{j}\alpha^j\beta^{n-j}$ 
	is the probability of having at least one male in a burden of $n$ parasites. 
	
	
	We will denote by $G$ to the pgf of the distribution of parasites per host $G_W$
	and  $G'(x)=\left.\frac{\partial G}{\partial s}\right|_{x}$.

	\subsection{Mating probability}
	From the above we obtain that the mating probability of a female, as the ratio between the mean number of fertilized females and the mean number of females in a host,
	\begin{equation*}
	\frac{\sum_{n\geq 0}\sum_{j=1}^{n-1}jp_n\binom{n}{j}\alpha^j\beta^{n-j}}
	{\sum_{n\geq 0}\sum_{j=0}^{n}jp_n\binom{n}{j}\alpha^j\beta^{n-j}}
	=\frac{\alpha  m -\alpha  G'(\alpha)}{\alpha m}
	\end{equation*}
	Therefore the probability of mating of a female that we will denote by $\phi$ is given by
	\begin{equation}\label{probrepro1}
	\phi=1-\frac{ G'(\alpha)}{m}
	\end{equation}

	\section{Mating probability and density-dependent fecundity}
	\subsection{Density-dependent fecundity}
	In population ecology, density-dependent processes
	(or density-dependent) occur when population growth rates are regulated by population density.
	
	In macroparasites life cycles, density-dependent processes can influence parasite fecundity, establishment and survival within host . 
	In the case of soil-transmitted helminths, there is a density-dependent fecundity in which the weight of females and their egg production rates decrease as the parasite burden on the host increases \cite{walker2009density,churcher2006density}.
	
	This negative density-dependence can be described mathematically by the negative exponential function
	\begin{equation}
	\lambda(n)=\lambda_0 \exp[-\gamma(n-1)]
	\end{equation} 
	where $\lambda(n)$ is the per capita female fecundity within a host with a parasite burden of size $n$,
	$\lambda_0$ is the intrinsic fecundity in absence of density-dependence effects and 
	$\gamma$ is the density-dependence intensity. 
	A study for Ascaris \textit{lumbricoides} is presented in \citep{hall2000geographical}.
	
	To simplify notation in rest of the text we will express the female fecundity by $\lambda(n)=\lambda_0 z^{n-1}$ where $z=e^{-\gamma}$.

	\subsection{Mean egg production per host}
	Due to the effects of density-dependent fecundity, the total egg production by females decreases as the parasite burden in host increases.
	Therefore, from the distribution of parasites per host, the mean egg production per host 
	is given by the expression
	\begin{equation}\label{egg}
	\sum_{n\geq 0}\sum_{j=0}^{n}j\lambda(n)p_n\binom{n}{j}\alpha^j\beta^{n-j}=\lambda_0\alpha G'(z)
	\end{equation}
	where $j\lambda(n)$ is the egg production of $j$ females and $p_n\binom{n}{j}\alpha^j\beta^{n-j}$ is the probability of having $j$ females, both cases within a host with $n$ parasites.
	

	\subsubsection{Mean fertilized egg production}
	For the fertilized egg production, we must consider only the fertilized females. Therefore the expression for the mean fertilized egg production is given by
	\begin{equation}\label{eggfecun}
	\sum_{n\geq 1}\sum_{j=1}^{n-1}j\lambda(n)p_n\binom{n}{j}\alpha^j\beta^{n-j}=
	\lambda_0 \alpha G'(z) \left[1-\frac{ G'(\alpha z)}{G'(z)}\right]  
	\end{equation}
	where, we obtain this expression by adding the female fecundity $\lambda(n)$ to what was developed in \eqref{eqhembrasfecun}.
	
	
	
	\subsubsection{Mean effective transmission contribution by female parasite}
	In mean-based deterministic population model  of parasite burden such as \cite{anderson1985helminth,anderson1992infectious,truscott2014modeling}, 
	is necessary to know the  term effective transmission contribution of female population to the reservoir (eggs or larvae)  \cite{churcher2005density,churcher2006density}.
	Using the results obtained in this work we can calculate this term denoted by $\psi$ as
	\begin{equation}\label{psi}
	\psi=\frac{\sum_{n\geq 0}\sum_{j=1}^{n}j\lambda(n)p_n\binom{n}{j}\alpha^j\beta^{n-j}}
	{\sum_{n\geq 0}\sum_{j=0}^{n}jp_n\binom{n}{j}\alpha^j\beta^{n-j}}
	=\frac{G'(z)}{m}   
	\end{equation}
	where the negative density-dependence function $\lambda(n)$ is redefined by $\lambda(n)/\lambda_0$.
	This allows the function $\lambda(n)$ to have a maximum value of 1
	and separate the density-independent term $\lambda_0$, from the density-dependent processes ($n$-dependent).

	
	\subsection{Mating probability and density-dependence effects}
	Due to the above, if we consider the ratio between the mean fertilized egg production and the mean egg production,
	we can obtain the fraction of the eggs that are fertilized by the male parasites,
	and therefore obtain the probability of fecundity of the eggs or mating probability of female parasites under the density-dependence effects
	\begin{equation}\label{probrepro2}
	\phi=1-\frac{G'(\alpha z)}{G'(z)} 
	\end{equation}
	If we consider this last expression \eqref{probrepro2} we notice that for the case where there is no density-dependence ($z \approx 1$) this expression is equivalent to expression \eqref{probrepro1}, therefore this is a generalization of the mating probability.
	
	On other hand, in case of mean-based deterministic model  of parasite burden, 
	we obtain that the contribution of fertilized egg production by mean parasite burden is modeled by the following expression in
	terms of functions $\psi$ and $\phi$
	\begin{equation}
	\lambda_0\alpha m \psi(m) \phi(m)= \lambda_0 \alpha G'(z) \left[1-\frac{ G'(\alpha z)}{G'(z)}\right] 	
	\end{equation}
	where we assume that $\psi$ and $\phi$ are functions of the mean $m$. 
	So we get the results of \citep{anderson1992infectious}.

	\section{Some examples}\label{sec:ejemplos}
	In this section we will consider the most common statistical models used to describe the distribution of parasites by host.
	\subsection{Poisson}
	For our first example we will consider a simple model for the distribution of parasites per host \cite{lahmar2001frequency}.
	In a Poisson model its probability mass function is of the form 
	%
	\begin{equation}
	\Pr(X=x)=\frac {\lambda ^{x}e^{-\lambda }}{x!},
	\end{equation}     
	where $\lambda$ is the mean parasite burden $m$ and its pgf is given by
	\begin{equation}
	\begin{split}
	G(s)&=e^{m(s-1)}\\
	\end{split}
	\end{equation}
	For this parasite distribution the 
	mean number of fertilized female parasites per host is given by
	$\alpha \lambda \left[1  -  e^{-m\beta} \right]$.
	On the other hand, the effective contribution of parasites to the transmission cycle is given by (see eq \eqref{psi})
	%
	\begin{equation}
	\psi=
	 e^{-m(1-z)}
	\end{equation}
	Another important term in parasite dynamics is the mating probability (general) $\phi$ which is given by (see eq \ref{probrepro2})
	%
	\begin{equation}
	\phi=
	1-e^{-mz \beta}
	\end{equation}
	This expression of $\phi$ results a generalization for the term mating probability obtained in the works \cite{anderson1992infectious,may1993biased,may1977togetherness}. 
	

	\subsection{Negative binomial}
	In the case of soil-transmitted helminths, works such as  \citep{bundy1987epidemiology,hoagland1978necator,seo1979frequency} show that the distribution of parasites by host can be described by a negative binomial model,
	\begin{equation}
	P(X=x)=\frac{\Gamma(k+x)}{\Gamma(x+1)\Gamma(k)}\left( \frac{k}{k+m}\right) ^k \left( \frac{m}{k+m}\right) ^x
	\end{equation}
	where $m$ is the mean parasite burden and $k$ is the inverse dispersion parameter of the parasites. Its pgf is given by
	\begin{equation}
	\begin{split}
	G(s)&=\left[ 1-\frac{m}{k}(s-1)\right] ^{-k}\\
	\end{split}
	\end{equation}
	Therefore the mean number of fertilized female parasites per host is given by the fraction
	$ 1-\left[ 1-\frac{m}{k}(\alpha-1)\right] ^{-(k+1)} $  of $\alpha m$. 
	Another important result is the expression for $\psi$, the effective contribution, which is given by (see eq. \eqref{psi})

	\begin{equation}\label{phibn}
	\psi=
	 \left[ 1-\frac{m}{k}(z-1)\right] ^{-(k+1)} 
	\end{equation}     
	Finally the mating probability, $\phi$, is given by (see eq. \eqref{probrepro2})
	\begin{equation} 
	\phi=
	1-\left[ \frac{ 1-\frac{m}{k}(\alpha z-1)}{1-\frac{m}{k}(z-1) }\right]  ^{-(k+1)} 
	\end{equation}
	This expression of $\phi$ results in a generalization for the mating probability obtained in works  \cite{anderson1992infectious,may1993biased,may1977togetherness}.
		
	\subsection{Zero-inflated and hurdle Models}
	Other frequently used models for event counting in parasites are the zero-inflated and hurdle models as mentioned in the works \citep{abdybekova2012frequency,crofton1971quantitative,denwood2008distribution,ziadinov2010frequency}.
	For a zero-inflated model, its probability mass function is of the form
	\begin{equation*}\label{zid}
	P(Y=y)= \left\{ \begin{array}{lc}
	\pi + (1-\pi)p_0 & y=0 \\
	\\ (1-\pi)p_y  & y\neq 0
	\end{array}
	\right.
	\end{equation*}
	where $p$ is the probability mass function of a distribution with no excess zeros.
	If $G$ is the pgf of the distribution with no excess zeros, the pgf of the zero-inflated distribution and its mean are of the form
	\begin{equation*}
	\begin{split}
	F(s)&=\pi+(1-\pi)G(s)\\
	m_F&=(1-\pi)m_G
	\end{split}
	\end{equation*}
	Then for this model the mean number of fertilized female parasites per host is given by
	
%
	\begin{equation*}
	\alpha F'(1) \left[1-  \frac{F'(\alpha )}{ F'(1)}\right]=\alpha (1-\pi) G'(1) 
	\left[1 - \frac{G'(\alpha)}{G'(1)}\right]    
	\end{equation*}
	Another important result is the expression for $\psi$, the mean contribution per female parasite, which is given by
	\begin{equation}\label{zipsi}
	\psi= \frac{F'(z)}{m_F}=  \frac{(1-\pi)G'(z)}{(1-\pi)m_G}=\frac{G'(z)}{m_G} 
	\end{equation}
	Finally the mating probability $\phi$ can be calculated by     
	\begin{equation}\label{ziphi}
	\phi=1-\frac{F'(\alpha z)}{F'(z)}=1-\frac{G'(\alpha z)}{G'(z)} 
	\end{equation}
	
	A hurdle model is a two-part model, 
	the first part $\pi$ which is the probability of attaining value zero, and the second part $1-\pi$ which is the probability of non-zero values. 
	The use of hurdle models is often motivated by an excess of zeros in the data, which is not sufficiently accounted for in more standard statistical models [5]. 
	For this model its probability mass function is given by
	\begin{equation*}\label{hd}
	P(Y=y)= \left\{ \begin{array}{lc}
	\pi & y=0 \\
	\\ (1-\pi)\frac{p(y)}{1-p_0}  & y\neq 0
	\end{array}
	\right.
	\end{equation*}
	Its pgf $H$ and its mean are of the form
	\begin{equation*}
	\begin{split}
	H(s)&=\pi+(1-\pi)\frac{G(s)-p_0}{1-p_0}\\
	m_H&=(1-\pi)\frac{m_G}{1-p_0}
	\end{split}
	\end{equation*}
	Therefore
	\begin{equation}
	\begin{split}
	\psi&= \frac{H'(z)}{m_H}= \frac{\rho G'(z;m_G)}{ m_H} =\frac{G'\left( z;\frac{m_H}{\rho}\right) }{\frac{m_H}{\rho}}\\
	\phi&=1-\frac{H'(\alpha z)}{H'(z)}=1-\frac{G'\left( \alpha z;\frac{m_H}{\rho}\right) }
	{G'\left( z;\frac{m_H}{\rho}\right) } 
	\end{split}
	\end{equation}
	where $\rho=\frac{1-\pi}{1-p_0}$.
	
	\subsubsection{Zero-inflated geometric}
\begin{figure}[h!]
		\centering
		\includegraphics[width=.99\linewidth]{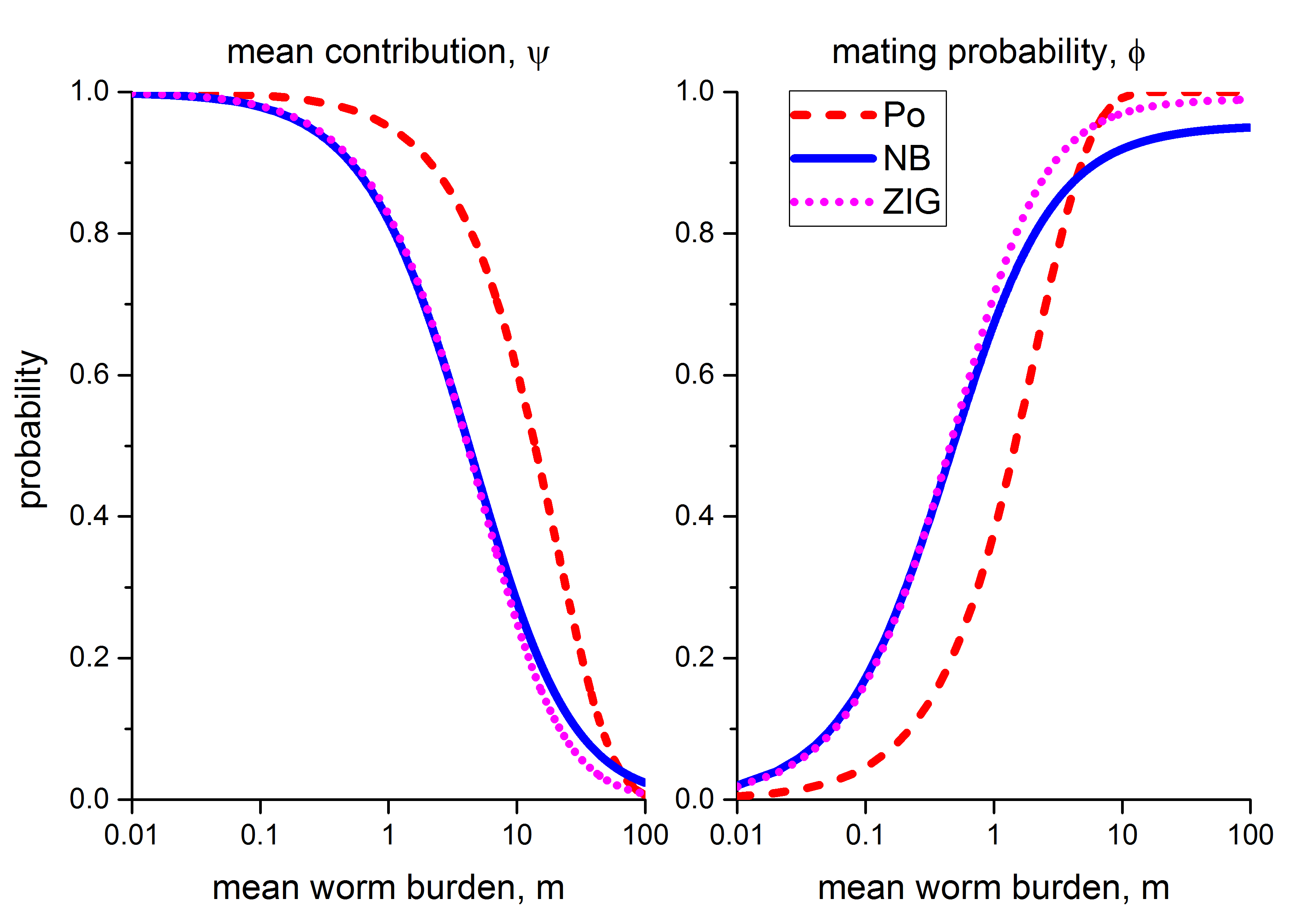}
		\caption{The mean effective contribution $\psi$ (left) and the mating probability $\phi$ (right) corresponding to Poisson (dash curve), negative binomial (solid curve) and zero-inflated geometric (dot curve) distributions. All as a function of the mean parasite burden $m$.}
		\label{fig:phi}
	\end{figure}

	The zero-inflated geometric distribution is given by
	\begin{equation}\label{zig}
	P(Y=y)= \left\{ \begin{array}{lc}
	\pi + (1-\pi)p & y=0 \\
	\\ (1-\pi)p q ^y  & y\neq 0
	\end{array}
	\right.
	\end{equation}
	Its pgf $G$ and its mean is given by
	\begin{equation}
	\begin{split}
	G(s)&=\left[ 1-\frac{1-p}{p}(s-1)\right] ^{-1}\\
	m_G&=\frac{1-p}{p}	
	\end{split}
	\end{equation}
	The effective contribution, $\psi$, is given by (see eq. \ref{zipsi})
	\begin{equation}\label{phibn}
	\psi=
	\left[ 1-\frac{m}{1-\pi}(z-1)\right] ^{-2} 
	\end{equation}     
	While the mating probability, $\phi$, is given by (see eq. \ref{ziphi})
	\begin{equation} 
	\phi=
	1-\left[ \frac{ 1-\frac{m}{1-\pi}(\alpha z-1)}{1-\frac{m}{1-\pi}(z-1) }\right]  ^{-2} 
	\end{equation}
	In Figure \ref{fig:phi} we show plots of the effective mean contribution $\psi$ and the matching probability $\phi$ for all the distributions discussed above. We consider the parameters $z=$0.95, $k=$0.3, $\pi=$0.5.

	\section{Independence in the variables $F$ and $M$}\label{sec:disindep}
	Let $W$ be the random variable count of the number of parasites in a host and $F$ , $M$ are the number of female and male parasites, respectively.
	In section  \ref{sec:distsexo} we assumed that the variables $F$ and $M$ were dependent. In this section we study the case in which these variables are independent, that is, $W$ ,$ F$ and $M$ verify the following properties
	\begin{equation}\label{independencia}
	\begin{split}
	W&=F+M\\
	G_W(s)&=G_F(s)G_M(s)
	\end{split}
	\end{equation}
	The independence of the variables $F$ and $M$ can occur when the parasites are acquired individually, as in case of hookworm parasites that can penetrate the skin of host \cite{paho2022,who2022}.
	Unlike the ingestion of eggs or larvae of helmint parasites where the host can acquire one or more parasites in the same event, entry through the skin is done individually for each parasite.
	
	We present all the expressions developed in the sections \ref{sec:distsexo} and \ref{sec:probapareamiento}. 
	\begin{itemize}
		\item Mean number of fertilized female parasites
		\begin{align}
		\alpha m \left[1-p_M(0) \right] 
		\end{align}
		
		\item Mating probability 
		\begin{align}
		1-p_M(0) 
		\end{align}
		
		\item Mean egg production per host
		\begin{equation}
		\lambda_0G_M(z)G'_F(z)
		\end{equation}
		
		\item Mean fertilized egg production
		\begin{equation}
		\lambda_0 G_M(z) G'_F(z)\left[ 1-\frac{p_M(0)}{G_M(z)}\right]
		\end{equation}
		
		\item Mean effective transmission contribution by female parasite
		\begin{align}
		\psi=\frac{G_M(z)G'_F(z)}{\alpha m}
		\end{align}
		
		\item Mating probability and density-dependence effects
		\begin{equation}
		\phi= 1-\frac{p_M(0)}{G_M(z)}
		\end{equation}
		
		\item Contribution of mean fertilized egg production for mean-based deterministic model  of parasite burden
		 \begin{equation}
		\lambda_0 \alpha m \psi(m) \phi(m)
		\end{equation}
	\end{itemize}

	\subsection{Some examples}
	In the examples presented here we intend that the variables $W$ , $F$ and $M$ correspond to the same statistical model.
	We work with some of the most popular distributions	used to model parasites. 
	Recall that we assume the sex ratios of female : male parasites to be $\alpha:\beta$, where $\alpha+\beta=1$.
	
	\subsubsection{Poisson}
	For the case where the distribution of parasites per host is Poisson with mean $\lambda$, that is, $W\sim \mathrm{Po}(\lambda)$. A solution for the independence of variables $F$ and $M$ are the following distributions
	\begin{equation*}
	F\sim \mathrm{Po}(\alpha \lambda) \qquad M\sim \mathrm{Po}(\beta \lambda)
	\end{equation*}
	\begin{align*}
	G_F(s)G_M(s)&=e^{\alpha\lambda(s-1)}e^{\beta\lambda(s-1)}\\
	&=e^{(\alpha+\beta)\lambda(s-1)}\\
	&=e^{\lambda(s-1)}\\
	&=G_{F+M}(s)\\
	&=G_W(s)
	\end{align*}
	Note that the pgf of $F$ and $M$ coincide with what was obtained in section \ref{sec:distsexo}, which shows the independence of these variables in that section.
	We show some of the expressions obtained in the previous section \ref{sec:distsexo} for case of independence between variables 
	\begin{itemize}
	\item Mean effective transmission contribution by female parasite
	\begin{align*}
	\psi=\frac{G_M(z)G'_F(z)}{G'_F(1)}=e^{-\lambda(1-z)}
	\end{align*}
	
	\item Mating probability and density-dependence effects
	\begin{equation*}
	\phi= 1-\frac{p_M(0)}{G_M(z)}=1-e^{-\lambda z \beta}
	\end{equation*}
	\end{itemize}
	Note that the expression for $\psi$ and $\phi$ are the same as those obtained in the section \ref{sec:ejemplos}.
	
	\subsubsection{Negative binomial}
	Assuming a negative binomial distribution, for the distribution of parasites per host, with mean $m$ and dispersion parameter $k$. A solution to problem (\ref{independencia}) is given by
	\begin{equation*}
	F\sim \mathrm{NB}(\alpha m,\alpha k) \qquad M\sim \mathrm{NB}(\beta m,\beta k)
	\end{equation*}   
	\begin{align*}
	G_F(s)G_M(s)&=\left[ 1-\frac{\alpha m}{\alpha k}(s-1)\right] ^{-\alpha k} \left[ 1-\frac{\beta m}{\beta k}(s-1)\right] ^{-\beta k}\\
	&=\left[ 1-\frac{m}{k}(s-1)\right] ^{-\alpha k-\beta k}\\
	&=\left[ 1- \frac{m}{k}(s-1) \right] ^{-k}\\
	&=G_{F+M}(s)\\
	&=G_W(s)
	\end{align*}
	For this case, the pgf of $F$ and $M$ are not equal to those obtained in section \ref{sec:distsexo}, since it was shown that the variables were not independent.
	We show some of the expressions obtained in the previous section \ref{sec:distsexo} for case of independence between variables 
	
\begin{figure}
		\centering
		\includegraphics[width=0.99\linewidth]{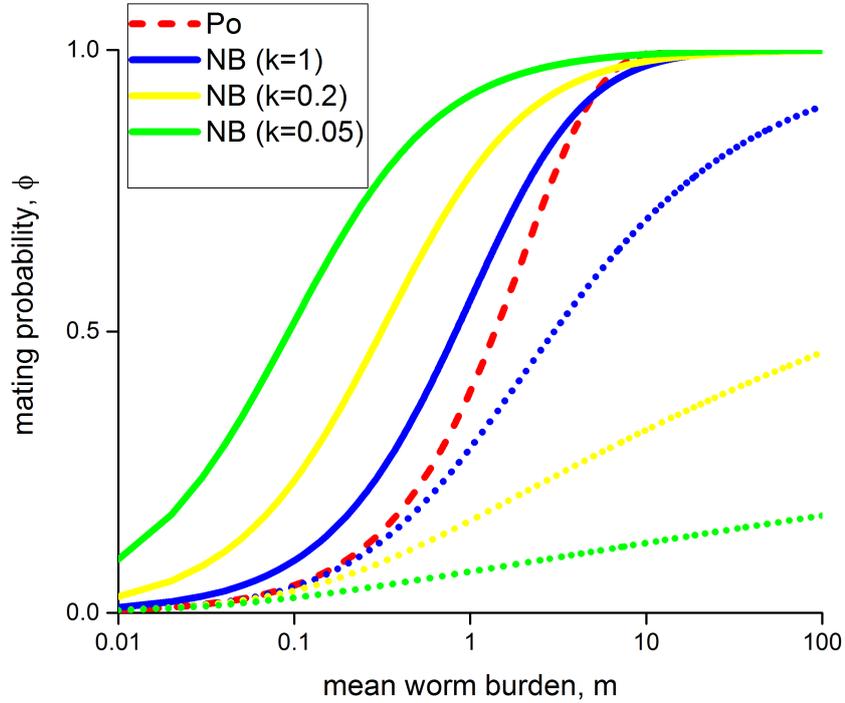}
		\caption{Mating probability as a function of mean parasite load. The dashed curve (red) corresponds to a Poisson distribution ($k\to \infty$). The solid and dotted curves correspond to a negative binomial distribution with joint or independent distribution by sex, respectively, where $k = 1$ (blue), $k =0.2$ (yellow) and $k =0.05$ (green).} 
		\label{fig:funphi}
	\end{figure}

	\begin{itemize}
		\item Mean effective transmission contribution by female parasite
		\begin{align}
		\psi=\frac{G_M(z)G'_F(z)}{\alpha m}=\left[ 1-\frac{m}{k}(z-1)\right] ^{-(k+1)}
		\end{align}
		
		\item Mating probability and density-dependence effects
		\begin{equation}
		\phi= 1-\frac{p_M(0)}{G_M(z)}=1-\left[ \frac{1+\frac{m}{k}}{1-\frac{m}{k}(z-1)}\right]^{-\beta k} 
		\end{equation}
	\end{itemize}
	Note that the expression $\psi$ is the same one obtained in the section \ref{sec:ejemplos}.
	However, this occurs with the mating probability $\phi$.
	In Figure \ref{fig:funphi} we show the behavior of the mating probability for both the case of a joint and independent sex distribution.

	\newpage
	\section{Discussion and Conclusions}

	Assuming an arbitrary model for distribution of parasites by host, we model the distributions of females and males.
	We model different reproductive variables of parasites such as mean number of fertilized female parasites, mean egg production, mating probability, mean fertilized egg production and mating probability and density-dependence effects.
	We show that these reproductive variables depend on independent nature of the $F$ and $M$ variables,  and  density-dependent fecundity of parasites.
	
	The reproductive expressions obtained in the examples of this work coincide with those obtained in\citep{leyton1968stochastic,may1993biased,may1977togetherness}.
	However, in these works, the effects of dense-dependent fertility on reproductive behavior of parasites are not considered.
	The expressions obtained are a generalization of expressions in \citep{leyton1968stochastic,may1993biased,may1977togetherness}.
	
	One of the main limitations of this work is that it only considers parasites with a polygamous mating system and we do not consider monogamous and hermaphroditic parasites.
	
	In conclusion, in this work we obtain a general expression for egg production and the mating probability of the parasites. 
	We show how these expressions depend on the sex distribution of the parasites and whether these distributions are considered joint or independent. 
	We also show that these expressions vary due to the effects of the density-dependence of the parasite.
	
%
%
%
%
%
%
%

	\section*{Aknowledgements}
	
	This work was partially supported by grant CIUNSA 2018-2467. JPA is a member of the CONICET. GML is a doctoral fellow of CONICET.
	
	\newpage
	\bibliographystyle{apa}
	\bibliography{biblio}	

\end{document}